\documentclass[sigconf]{acmart-me}

\usepackage{booktabs} 
\usepackage{url}
\usepackage{color}
\usepackage{enumitem}
\usepackage{todonotes}

\usepackage{MnSymbol,wasysym}

\setcopyright{rightsretained}

\acmDOI{}

\acmISBN{}

\acmConference[MediaEval'20]{Multimedia Evaluation Workshop}{December 2020}{Online} 
\acmYear{2020}
\copyrightyear{}

\acmPrice{}

\begin{document}
\title{Overview of MediaEval 2020 Predicting Media Memorability Task: What Makes a Video Memorable?}


\author{Alba G. Seco de Herrera\textsuperscript{1}, Rukiye Savran Kiziltepe\textsuperscript{1}, Jon Chamberlain\textsuperscript{1}, Mihai Gabriel Constantin\textsuperscript{2}, Claire-Hélène Demarty\textsuperscript{3}, Faiyaz Doctor\textsuperscript{1},  Bogdan Ionescu\textsuperscript{2}, Alan F. Smeaton\textsuperscript{4}}
\affiliation{\textsuperscript{1}University of Essex, UK\\ \textsuperscript{2}University Politehnica of Bucharest, Romania\\ \textsuperscript{3}InterDigital, R\&I, France\\ \textsuperscript{4}Dublin City University, Ireland.}
\email{alba.garcia@essex.ac.uk}

%
%
%
%
%

\renewcommand{\shortauthors}{A.G. Seco de Herrera et al.}
\renewcommand{\shorttitle}{Predicting Media Memorability}

\begin{abstract}
This paper describes the MediaEval 2020 \textit{Predicting Media Memorability} task. After first being proposed at MediaEval 2018, the \textit{Predicting Media Memorability} task is in its 3rd edition this year, as the prediction of  short-term and long-term video memorability (VM) remains a challenging task. In 2020, the format remained the same as in previous editions. This year the videos are a subset of the TRECVid 2019 Video-to-Text dataset, containing more action rich video content as compared with the 2019 task.
In this paper a description of some aspects of this task is provided, including its main characteristics, a description of the collection, the ground truth dataset, evaluation metrics and the requirements for participants' run submissions.
\end{abstract}

%
%
%
%
%


\maketitle
\section{Introduction}
\label{sc:intro}
%
Media platforms such as social networks, media advertisements, information retrieval and recommendation systems deal with exponential growth. Enhancing the relevance of multimedia occurrences in our everyday lives requires new ways to organise – in particular, to retrieve – digital content. Like other video metrics of importance, such as aesthetics or interestingness, memorability can be regarded as a useful aspect to help make a choice between competing videos. This is even truer when one considers  specific use cases of creating commercials or educational content. Because the impact of different multimedia content, images or videos, on human memory is unequal, the capability of predicting the memorability of a given piece of video content is of high importance for professionals in the field of advertising and other fields. Beyond advertising, other applications, such as film-making, education, content retrieval, etc., may also be influenced by this task.

The \textit{Predicting Media Memorability} task addresses this problem. The task is part of the MediaEval benchmark and, following the success of previous editions~\cite{CID2019,CDD2018}, creates a common benchmarking protocol and provides a ground truth dataset for short-term and long-term memorability using common definitions. 

%
\section{Related Work}
\label{sc:work}
%
The computational understanding of video memorability follows on from the study of image memorability prediction, which has attracted increasing attention since the seminal work of Isola et al.~\cite{IXP2013}. Models have achieved very good results at predicting image memorability~\cite{KRT2015,SDG2018} and we have recently started to see the use of techniques like style transfer to improve image memorability~\cite{10.1145/3311781} thus illustrating that we have now moved from just measuring memorability, to using memorability as an evaluation metric. 

In contrast, research on visual memorability (VM) from a computer science point of view is in its early stage. Recently we have seen other work on video memorability~\cite{10.1007/978-3-030-58517-4_14} with a particular focus on short term, but the scarcity of studies on VM can be explained by several reasons. Firstly, there is no publicly available data set to train and test models, though the VideoMem~\cite{SSS2017} and the Memento10k~\cite{10.1007/978-3-030-58517-4_14} datasets are recent additions. The second point, closely related to the previous one, is the lack of a common definition for VM. Regarding modelling, previous attempts at predicting VM~\cite{SSS2017,CDD2019} have highlighted several features which contribute to the prediction of VM, such as semantic, saliency and colour features, but the work is far from complete and our capacity to propose effective computational models will help to meet the challenge of VM prediction. 

The goal of this task is to participate in the harmonisation and the advancement of this emerging multimedia field. Furthermore, in contrast to previous work on image memorability prediction, where memorability was measured a few minutes after memorisation, we propose a dataset with longer term memorability annotations. We expect the predictions of the models trained on this data to be more representative of long-term memory, which is used preferably in numerous applications.

\section{Task Description}
\label{sc:task}
%
The \textit{Predicting Media Memorability} task requires participants to automatically predict memorability scores for short form videos, that reflect the probability for a video to be remembered. Participants were provided with a dataset of videos with short-term and long-term memorability annotations, related information, and pre-extracted state-of-the-art visual features. 
Therefore, two subtasks were proposed to participants:
\begin{itemize}
    \item \textbf{Short-term VM prediction} - scores were
measured a few minutes after the memorisation process;
    \item \textbf{Long-term VM prediction} - scores were
measured 24-72 hours after the memorisation process.
\end{itemize}
\section{Collection}
\label{sc:collection}
\begin{figure}
\includegraphics[width=0.45\textwidth]{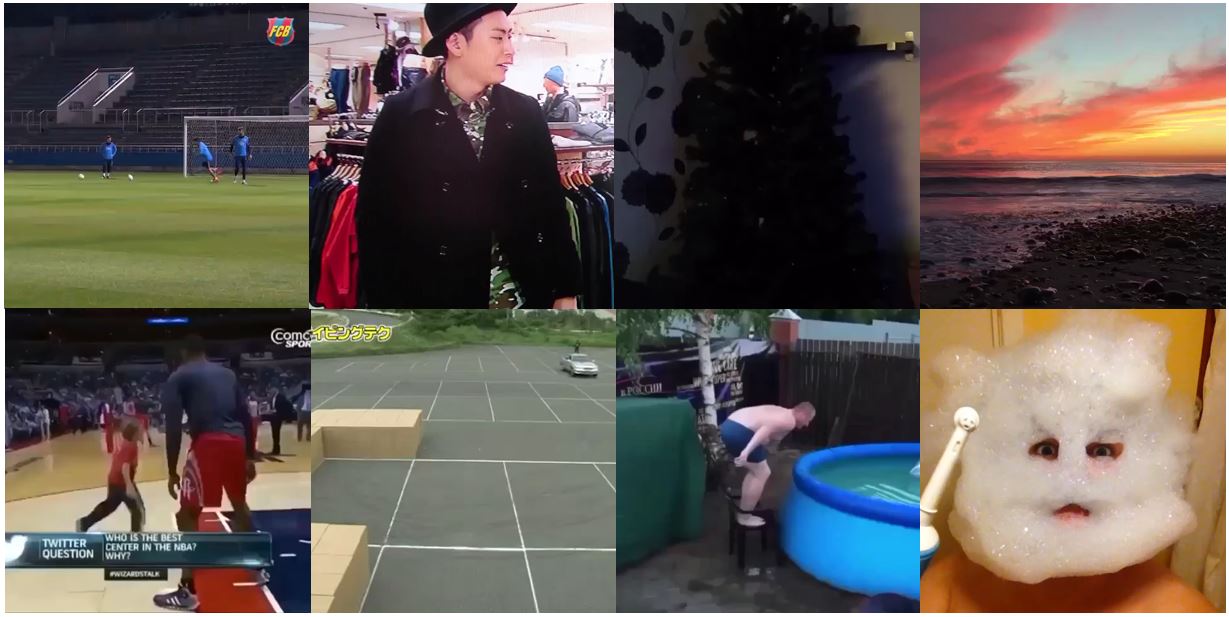}
\caption{A sample of frames of the videos in the \textit{TRECVid 2019 Video-to-Text dataset.}}
\label{fg:videos}
\end{figure}

The dataset is composed of a subset of short videos selected  from the TRECVid 2019 Video-to-Text dataset~\cite{ABC2019} (see Figure~\ref{fg:videos}). These videos are shared under Creative Commons licenses that allow their redistribution. The TRECVid videos have much more action happening in them compared with those in the 2019 VM task, and thus they correspond to more generic use cases. 

Each video consists of a coherent unit in terms of meaning and is associated with two scores of memorability that refer to its probability to be remembered after two different time durations of memory retention. 
A set of pre-extracted features are also distributed:
\begin{itemize}
    \item image-level features: AlexNetFC7~\cite{krizhevsky2012imagenet}, HOG~\cite{dalal2005histograms}, HSVHist, RGBHist, LBP~\cite{he1990texture}, VGGFC7~\cite{simonyan2014very};
    \item video-level feature: C3D~\cite{tran2015learning}. 
\end{itemize}

\noindent 
The image-level features were extracted from 3 frames for each video: the first, the middle and the last frame. In addition, each TRECVid video is accompanied by two textual captions describing the activity. Additional information on the annotation was also provided to allow further investigation of the user interaction for memorability. Hence, the annotations collected from participants were provided including the first appearance position and the second appearance position of each target video along with the response time of the user and the key pressed when watching each video.

The TRECVid 2019 Video-to-Text dataset~\cite{ABC2019} contains 6,000 videos. In 2020, three subsets were distributed as part of the MediaEval Predicting Media Memorability task.
The training set contained 590 videos, the development set 410 videos and the test set 500 videos. Each video was annotated by at least  16 annotators for  their short term memorability. However, there are fewer long term annotations.
%
%
%

Similar to previous editions of the task~\cite{CDD2018,CID2019}, memorability has been measured  using recognition tests, i.e., through an objective measure, a few minutes after the memorisation of the videos (short term), and then 24 to 72 hours later (long term). The ground truth dataset was collected by using a video memorability game protocol proposed by Cohendet et al.~\cite{CDD2019}. Two versions of the memorability game were published. One  was published on Amazon Mechanical Turk (AMT) and another one was issued for general use with following three language options: English, Spanish and Turkish.

In the game of video memorability, participants are expected to watch 180 and 120 videos in short-term and long-term memorisation steps, respectively. The task is basically to press the space bar once the participants recognise a previously seen video, which enables to determine videos recognised and not recognised  by them. In the first step of the game, 40 target videos are repeated after a few minutes to collect short-term memorability labels. As for filler videos in the first step, 60 non-vigilance filler videos are displayed once. 20 vigilance filler videos are repeated after a few seconds to check participants' attention to the task. After 24 hours to 72 hours, the same participants are expected to attend the second step for collecting long-term memorability labels. This time, 40 target videos chosen randomly from among non-vigilance fillers of the first step and 80 fillers selected randomly from new videos are displayed to measure long-term memorability scores for those target videos. Both short-term and long-term memorability scores are calculated as the percentage of correct recognition for each video by the participants. Relevant screenshots and label collection procedures are demonstrated on the MediaEval task web page \cite{benchmark_2020}.

\section{Submission and Evaluation}
\label{sc:run}
%
As in previous editions of the task, each team is required to predict both short and long term memorability. In total, 10 runs can be submitted, 5 for each.
For the two required runs, all information can be used in the development of the system, meaning provided features, ground truth data, video sample titles, features extracted from the visual content and even external data. The only exception, in
this case, is that the required short-term and long-term memorability runs must not use each other's score annotations. For the rest of the runs, a maximum of 4 per subtask, everything is permitted, including using cross-annotations between the subtasks.
%

The outputs of the prediction models – i.e., the predicted memorability scores for the videos – will be compared with ground truth memorability scores using classic evaluation metrics (e.g., Spearman’s rank correlation).
%
\section{Discussion and Outlook}
\label{sc:discussion}
%
In this paper we presented the third edition of the Predicting Media Memorability at the MediaEval 2020 Benchmarking initiative. The task provides a framework that allows a comparative study of different state of the art Machine Learning approaches aiming to predict short and long-term memorability. A collection of videos is provided as well as memorability annotations and a common evaluation metric. In addition, related information has been provided to help participants in developing their approaches.
Details regarding the methods employed by participants and their results can be found in the proceedings of the 2020 MediaEval workshop\footnote{See CEUR Workshop Proceedings (CEUR-WS.org).}.
\begin{acks}
This work was part-funded by NIST Award No. 60NANB19D155, by Science Foundation Ireland under grant number SFI/12/RC/2289\_P2 and under project AI4Media, A European Excellence
Centre for Media, Society and Democracy, H2020 ICT-48-2020, grant 951911.
\end{acks}

\bibliographystyle{ACM-Reference-Format}
\def\bibfont{\small} 
\bibliography{references} 

\end{document}